\documentclass[aps,twocolumn,superscriptaddress,floatfix,nofootinbib,groupedaddress]{revtex4-1}
\usepackage{graphicx,amsmath,amssymb,verbatim,color}
\usepackage{booktabs}
\usepackage{comment}
\usepackage[utf8]{inputenc}
\usepackage[colorlinks=true,citecolor=blue,linkcolor=blue,urlcolor=blue, backref=false,pdfborder={0 0 0}]{hyperref}

\newcommand{\be}{\begin{equation}\begin{gathered}}
\newcommand{\ee}{\end{gathered}\end{equation}} 
\newcommand{\barr}{\begin{eqnarray}}
\newcommand{\earr}{\end{eqnarray}} 

\newcommand{\rme}{\textrm{e}}
\newcommand{\nh}{n_{\rm H}}
\newcommand{\bs}{\boldsymbol}
\newcommand{\hyrec}{\textsc{hyrec}}
\newcommand{\full}{\textsc{full}}
\newcommand{\recfast}{\textsc{recfast}}

\newcommand{\apjl}{ApJL}
\newcommand{\apjs}{ApJS}
\newcommand{\mnras}{MNRAS}
\newcommand{\jcap}{JCAP}
\newcommand{\aap}{Astron. Astrophys.}

\begin{document}

\title{\texttt{HYREC-2}: a highly accurate sub-millisecond recombination code}
\author{Nanoom Lee}
\email{nanoom.lee@nyu.edu}
\affiliation{Center for Cosmology and Particle Physics, Department of Physics, New York University, New York, NY 10003} 
\author{Yacine Ali-Ha\"imoud}
\email{yah2@nyu.edu}
\affiliation{Center for Cosmology and Particle Physics, Department of Physics, New York University, New York, NY 10003} 
\date{\today} 
\begin{abstract}
We present the new recombination code \hyrec-2, holding the same accuracy as the state-of-the-art codes \hyrec~and \textsc{cosmorec} and, at the same time, surpassing the computational speed of the code \recfast~commonly used for CMB-anisotropy data analyses. \hyrec-2 ~is based on an effective 4-level atom model, accounting exactly for the non-equilibrium of highly excited states of hydrogen, and very accurately for radiative transfer effects with a correction to the Lyman-$\alpha$ escape rate. The latter is computed with the original \hyrec, and tabulated, as a function of temperature, along with its derivatives with respect to the relevant cosmological parameters. This enables the code to keep the same accuracy as the original \hyrec~over the full 99.7\% confidence region of cosmological parameters currently allowed by \emph{Planck}, while running in under one millisecond on a standard laptop. Our code leads to no noticeable bias in any cosmological parameters even in the case of an ideal cosmic-variance limited experiment up to $\ell = 5000$. Beyond CMB anisotropy calculations, \hyrec-2~will be a useful tool to compute various observables that depend on the recombination and thermal history, such as the recombination spectrum or the 21-cm signal.
\end{abstract}
\maketitle

\section{Introduction}

The recombination history of the Universe is a key part of the physics of Cosmic Microwave Background (CMB) anisotropies, the epoch of the dark ages leading to the formation of the first stars, as well as the formation of cosmic structure. When exactly free electrons got bound in the first helium and hydrogen atoms determines, first, the epoch of photon last scattering, thus the sound horizon. This scale is imprinted into CMB power spectra and the correlation function of galaxies, and serves as a standard ruler \cite{Eisenstein_98}. The abundance of free electrons also sets the photon diffusion scale, hence the damping of small-scale CMB anisotropies \cite{Silk_68}. Lastly, the free-electron fraction determines the epochs of kinematic and kinetic decoupling of baryons from photons, hence the thermal history of the gas, as well as the formation of the first stars and structures.

The basic physics of hydrogen recombination were laid out in the late sixties in the seminal works of Peebles \cite{Peebles:1968ja} and Zeldovich, Kurt, and Sunyaev \cite{Zeldovich_69}. Their simple but physically accurate effective 3-level model was largely unchanged until the late nineties (see Ref.~\cite{Hu_95} for an overview of recombination studies till then), except for improvements in the atomic-physics calculations of case-B recombination coefficients \cite{Pequignot_91}. In 1999, motivated by the approval of the WMAP \cite{Bennett_03} and \emph{Planck} \cite{Planck_06} satellites, Seager, Sasselov \& Scott conducted the first modern, detailed recombination calculation \cite{Seager:1999km}, explicitly accounting for the non-equilibrium of highly excited hydrogen energy levels (but assuming equilibrium amongst angular momentum substates). They found that the result of their 300-level calculation could be accurately reproduced by an effective 3-level atom model, with the case-B recombination coefficient multiplied by a ``fudge factor" $F = 1.14$ \cite{Seager:1999bc}. This model was implemented in the code \recfast, which was used for cosmological analyses of WMAP data \cite{Hinshaw_13}, for which it was sufficiently accurate. 

It was realized in the mid-2000's that the \recfast~model for hydrogen recombination would not be sufficiently accurate for the analysis of \emph{Planck} data, as it neglected a variety of physical effects that matter at the required sub-percent level of accuracy (see Ref.~\cite{Rubino-Martin_10} for an overview of progress by the end of that decade). On the one hand, the angular momentum substates of the excited states of hydrogen are out of equilibrium, which leads to an overall slow-down of recombination \cite{Rubino-Martin_06, Chluba_07, Grin_10, Chluba_10}. On the other hand, a variety of radiative transfer effects have to be accounted for, such as feedback from higher-order Lyman transitions, frequency diffusion due to resonant scattering, and two photon transition from higher levels \cite{Chluba:2007yp,Kholupenko:2009my,Chluba:2005uz,Dubrovich:2005fc,Chluba:2008vn,Grachev:2008xj, Hirata:2009qy,Chluba:2009gv}.

While conceptually straightforward, the inclusion of hydrogen's angular momentum substates presented a considerable computational challenge with the standard multilevel method. Indeed, Refs.~\cite{Grin_10, Chluba_10} showed that a recombination history converged at the level needed for \emph{Planck} requires accounting for at least 100 shells of hydrogen energy levels, corresponding to about 5000 separate states. The standard multilevel approach required solving large linear systems at each time step, and even the fastest codes took several hours per recombination history on a standard single-processor machine \cite{Chluba_10}. This aspect of the recombination problem was solved conclusively a decade ago in Ref.~\cite{AliHaimoud:2010ab}, where it was shown that the non-equilibrium dynamics of the excited states can be accounted for \emph{exactly} with an effective few-level atom model (in practice, 4 levels are enough), with effective recombination coefficients to the lowest excited states accounting for intermediate transitions through the highly excited states (see also \cite{Burgin_09, Burgin_10} for an independent discovery of the method). In contrast with \recfast's fudged case-B coefficient, these effective rates are exact, temperature-dependent atomic physics functions. Once this computational hurdle was cleared, efficient methods to solve the radiative transfer problem were developed shortly after, leading to the fast state-of-the-art public recombination codes \hyrec~\cite{AliHaimoud:2010dx} and \textsc{cosmorec} \cite{Chluba:2010ca}, in excellent agreement with one another despite their different radiative transfer algorithms. The residual theoretical uncertainty of these codes is estimated at the level of a few times 10$^{-4}$ during hydrogen recombination, due to the neglect of subtle radiative transfer effects \cite{Ali-Haimoud_10c} and collisional transitions \cite{Chluba_10}, whose rates are uncertain.

The accuracy requirement for helium recombination is not as stringent as it is for hydrogen, given that it recombines well before the time at which most CMB photons last scattered. Still, a variety of important radiative transfer effects must be accounted for at the level of accuracy required for \emph{Planck}, such as the photoionization of neutral hydrogen atoms by resonant 584 \AA~photons and the emission of intercombination-line photons at 591 \AA~\cite{Switzer_08a, Hirata_08, Switzer_08b, Rubino-Martin_08, Kholupenko_08}. These effects are included numerically in \textsc{cosmorec} and through fast analytic approximations in \hyrec, accurate within 0.3\%, which is sufficient for \emph{Planck}. In the rest of this paper, we focus on hydrogen recombination. We defer the task of extending our approach to helium to future work.

Both \hyrec~and \textsc{cosmorec} are interfaced with the commonly used Boltzmann codes \textsc{camb} \cite{Lewis:1999bs, Howlett:2012mh} and \textsc{class} \cite{Lesgourgues:2011re}, and are able to compute a recombination history in about half a second on a standard laptop. Still, the default code for the cosmological analysis of \emph{Planck} data has remained \recfast~\cite{Aghanim:2018eyx}, further modified to approximately reproduce the output of \hyrec~and \textsc{cosmorec}. The non-equilibrium of angular momentum substates is approximately accounted for by lowering the case-B coefficient fudge factor from 1.14 to 1.125. Radiative transfer physics are approximately mimicked by introducing a double-Gaussian ``fudge function", correcting the net decay rate in the Lyman-$\alpha$ line\footnote{To our knowledge, there is no publication describing how the form of the fudge function and the best-fit parameters were determined, nor quantifying the residual error and its impact on cosmological parameter estimation for future experiments.}. The advantage of this re-fudged \recfast~over \hyrec~and \textsc{cosmorec} remains speed: by not explicitly solving a radiative transfer problem, \recfast~computes a recombination history in about 0.03 second on a standard laptop. The recombination calculation is not parallelizable, in contrast with the computation of the transfer functions of independent Fourier modes in a Boltzmann code. Therefore, the additional time spent by \hyrec~and \textsc{cosmorec} can be the bottleneck of CMB anisotropy calculations, which may explain the choice of using \recfast~over its more modern, accurate and versatile counterparts for \emph{Planck} analyses. 

As we confirm in this work, the re-fudged \recfast~is sufficiently accurate for the analysis of \emph{Planck} data, in the sense that it leads to biases in cosmological paramaters much smaller than their statistical uncertainties. However, \emph{Planck} is not the final CMB-anisotropy mission: the Simons Observatory \cite{SO_19} and CMB stage-IV \cite{Abazajian:2016yjj} which are ground-based surveys, will have more than 10 times better sensitivity with a comparable sky coverage; the proposed CORE satellite \cite{Core} will have 10-30 times better sensitivity with full sky coverage. It is unclear whether \recfast~is sufficiently accurate for future CMB missions, nor whether simple additional fudges would be sufficient.

In this paper, we describe the new recombination code \hyrec-2\footnote{\hyrec-2~is available at \href{https://github.com/nanoomlee/HYREC-2}{https://github.com/nanoomlee/HYREC-2}}, able to compute a recombination history with virtually the same accuracy as the original \hyrec, in under 1 millisecond on a standard laptop. Our new code therefore surpasses \recfast~in both accuracy and speed, and ought to become the standard tool for the analysis of future CMB-anisotropy data. \hyrec-2~is based on an effective 4-level atom model, accounting exactly for the non-equilibrium of excited states of hydrogen \cite{AliHaimoud:2010ab}, hence accurately capturing the low-redshift tail of recombination, without requiring any fudge factors. Radiative transfer effects are accounted for with a redshift- and cosmology-dependent correction to the Lyman-$\alpha$ net decay rate, exact up to errors quadratic in the deviations of cosmological parameters away from the \emph{Planck} 2018 best-fit cosmology \cite{Aghanim:2018eyx}. We check the accuracy of our new code extensively by sampling the full 99.7\% confidence region of the \emph{Planck} posterior distribution (assuming a Gaussian distribution), and verifying that the tiny difference with \hyrec~leads to negligible biases, even for futuristic CMB missions for which \recfast~would be insufficiently accurate.

The rest of this paper is organized as follows. In Section \ref{sec:models}, we briefly review hydrogen recombination physics and lay out the exact effective 4-level atom equations. In Section \ref{sec:swift}, we describe \hyrec-2, and quantify its accuracy in Section \ref{sec:results}. We conclude in Section \ref{sec:conclusion}. Appendix \ref{app:Delta} provides explicit equations for the correction functions used in \hyrec-2. In Appendix \ref{app:recfast}, we provide the equations used in \recfast~in our notation, for completeness and ease of comparison.

\section{Hydrogen recombination physics}\label{sec:models}

\subsection{Recombination phenomenology}

The basic phenomenology of hydrogen recombination has been well understood since the late sixties, with the seminal works of Peebles \cite{Peebles:1968ja} and Zeldovich, Kurt, and Sunyaev \cite{Zeldovich_69}. We briefly summarize the essential physics here (see e.g.~\cite{AliHaimoud_thesis} for more details) and introduce some of the notation along the way. 

Direct recombinations to the ground state are ineffective, as the emitted photons almost certainly reionize another hydrogen atom. Therefore, recombinations proceed through the excited states, with principal quantum number $n>1$. Once an electron and a proton bind together, the newly formed excited hydrogen atom undergoes rapid transitions to other excited states, and eventually either gets photoionized again by thermal CMB photons, or reaches the lowest excited state $n = 2$, with angular momentum substates $2s$ and $2p$. The net flow of electrons to the $2s$ and $2p$ states is described by effective recombination coefficients $\mathcal{A}_{2s}(T_m, T_r), \mathcal{A}_{2p}(T_m, T_r)$, which are pure atomic physics functions depending on matter and radiation temperatures only \cite{AliHaimoud:2010ab} (see also \cite{Burgin_09, Burgin_10}). 

Once in one of the $n = 2$ states, a hydrogen atom has three possible fates. First, it may get directly or indirectly photoionized by thermal CMB photons, with effective photoionization rates $\mathcal{B}_{2s}(T_r)$, $\mathcal{B}_{2p}(T_r)$, depending on the radiation temperature only \cite{AliHaimoud:2010ab} and related to the effective recombination coefficients through detailed balance relations. Second, it may indirectly transition to the other $n = 2$ state through intermediate transitions to higher excited states; the effective transition rates $\mathcal{R}_{2s, 2p}(T_r) = 3 \mathcal{R}_{2p, 2s}(T_r)$ are also pure functions of atomic physics which only depend on the radiation temperature \cite{AliHaimoud:2010ab}. Last, but not least, it may decay to the ground state. From the $2p$ state, hydrogen can efficiently decay to the ground state through the allowed Lyman-$\alpha$ transition; this resonant line is highly optically thick, however, and the vast majority of Lyman-$\alpha$ photons end up re-exciting another ground-state atom. The net transition rate in the Lyman-$\alpha$ line is therefore, \emph{approximately} the rate at which photons redshift out of the resonance due to cosmological expansion. For a sub-percent accuracy, one must calculate the net decay rate by solving the radiative transfer equation for resonant photons, accounting for feedback from higher-order Lyman lines \cite{Chluba:2007yp,Kholupenko:2009my}, two-photon transitions from higher levels \cite{Chluba:2005uz,Dubrovich:2005fc}, time dependent effects \cite{Chluba:2008vn}, and frequency diffusion in Lyman-$\alpha$ \cite{Grachev:2008xj, Hirata:2009qy,Chluba:2009gv}. From the $2s$ state, hydrogen may directly decay to the ground state through a ``forbidden" two-photon transition. While this transition is optically thin, the net decay rate is affected by the re-absorption of non-thermal photons redshifting out of the Lyman-$\alpha$ resonance \cite{Kholupenko:2006jm}, and the two-photon transition rate must be computed within the radiative transfer calculation. We denote by $\dot{x}_{2s}|_{1s}$,  $\dot{x}_{2s}|_{1s}$ the net rates of change of the fractional populations of n = 2 excited states through transitions to the ground state.

At $z \gtrsim 800$, atoms reaching the $n = 2$ states are more likely to be photoionized than reaching the ground state. Transitions to the ground state are thus the bottleneck of the recombination process at high redshifts, and as a consequence any error on the rates $\dot{x}_{2s}|_{1s}$, $\dot{x}_{2p}|_{1s}$ directly translates to an error on the overall recombination rate. At $z \lesssim 800$, atoms that reach $n = 2$ almost certainly decay to the ground state before being photoionized, and the recombination dynamics is controlled by the rate of recombinations to excited states, rather than decays to the ground state.

\subsection{General recombination equations}
Once helium has fully recombined, the following equation governs the evolution of the free-electron fraction $x_e$:
\be
\dot{x}_e = \sum_{\ell = s, p} \left( x_{2 \ell} \mathcal{B}_{2 \ell} - n_{\rm H} x_e^2 \mathcal{A}_{2 \ell} \right), \label{eq:xe-dot-exact}
\ee
where $n_{\rm H}$ is the total hydrogen density (both neutral and ionized), and $x_{2s}$, $x_{2p}$ are the fractional abundances of hydrogen in the first excited states. They are in turn determined by solving the coupled quasi-steady-state rate equations
\barr
0 \approx \dot{x}_{2 \ell} &=& n_{\rm H} x_e^2 \mathcal{A}_{2 \ell} - x_{2 \ell} \mathcal{B}_{2 \ell} + x_{2\ell'} \mathcal{R}_{2\ell', 2\ell} - x_{2\ell} \mathcal{R}_{2\ell, 2\ell'} \nonumber\\
&+& \dot{x}_{2\ell}|_{1s}, \label{eq:steady-state}
\earr
where $\ell' = p$ if $\ell = s$ and vice-versa.

The state-of-the-art recombination codes \hyrec~\cite{AliHaimoud:2010ab,AliHaimoud:2010dx} and \textsc{cosmorec} \cite{Chluba:2010ca} accurately compute the rates $\dot{x}_{2\ell}|_{1s}$, hence the populations of the first excited states $x_{2 \ell}$ and the net recombination rate from Eq.~\eqref{eq:xe-dot-exact} in their default modes (in \hyrec, the default mode is the ``\full" mode). They do so by solving the time-dependent radiative transfer equation, with different numerical algorithms, and agree with each other within their quoted uncertainty of a few parts in $10^4$. While they are much faster than the previous generation of recombination codes, the $\sim 1$ second per recombination history can become the bottleneck of CMB power spectra calculations, as it is not parallelizable.     

\subsection{Exact effective four-level equations} \label{four-level}

We may always formally write the net decay rate from $2\ell$ to the ground state in the form 
\barr
\dot{x}_{2 \ell}|_{1s} = - \mathcal{R}_{2\ell, 1s}(z)\left( x_{2 \ell}  - g_{2\ell} x_{1s} \rme^{- E_{21}/T_r} \right), \label{eq:xi_1s}
\earr
where $g_{2s} = 1$ and $g_{2p} = 3$ are the statistical weights of the $2s$ and $2p$ states, and $E_{21} \approx 10.2$ eV is the energy difference between the first excited state and the ground state. In contrast with the effective rates $\mathcal{A}_{2 \ell}$, $\mathcal{B}_{2 \ell}$, and $\mathcal{R}_{2 \ell,2 \ell'}$, the rates $\mathcal{R}_{2\ell, 1s}(z)$ are \emph{not} just functions of temperature: they depend on cosmological parameters through the expansion rate and hydrogen abundance, as well as on the full recombination history up to redshift $z$, due to the time-dependent nature of radiative transfer. 

Inserting Eq.~\eqref{eq:xi_1s} into the steady-state equations, one can find explicit expressions for $x_{2 \ell}$:
\barr
x_{2\ell} - g_{2 \ell} x_{1s} \rme^{- E_{21}/T_r} = \frac{\nh x_e^2 \mathcal{A}_{2 \ell} - g_{2 \ell} x_{1s} \rme^{- E_{21}/T_r} \mathcal{B}_{2 \ell}}{\Gamma_{2 \ell} - \mathcal{R}_{2 \ell, 2 \ell'}\mathcal{R}_{2 \ell', 2 \ell}/\Gamma_{2 \ell'}}\nonumber\\
+ \frac{\mathcal{R}_{2 \ell', 2 \ell}}{\Gamma_{2 \ell'}}\times  \frac{\nh x_e^2 \mathcal{A}_{2 \ell'} - g_{2 \ell'} x_{1s} \rme^{- E_{21}/T_r} \mathcal{B}_{2 \ell'}}{\Gamma_{2 \ell'} - \mathcal{R}_{2 \ell', 2 \ell}\mathcal{R}_{2 \ell, 2 \ell'}/\Gamma_{2 \ell}},~~~~\label{eq:x2l}
\earr
where $\Gamma_{2\ell}$ is the effective inverse lifetime of $2\ell$:
\be
\Gamma_{2 \ell}\equiv \mathcal{B}_{2\ell}+\mathcal{R}_{2 \ell, 2 \ell'}+ \mathcal{R}_{2 \ell, 1s},
\label{Gamma}
\ee
Inserting these expressions into Eq.~\eqref{eq:xe-dot-exact}, one finds \cite{AliHaimoud:2010ab}
\barr
\dot{x}_e &=& -\sum_{\ell = s, p} C_{2\ell}\left(\nh x_e^2 \mathcal{A}_{2\ell} - g_{2 \ell} x_{1s} \rme^{-E_{21}/T_r}~\mathcal{B}_{2 \ell} \right),\label{4level xe}
\earr
where the $C_{2 \ell}$-factors are given by
\barr
C_{2\ell}&\equiv& \frac{\mathcal{R}_{2 \ell, 1s} + \mathcal{R}_{2 \ell,2 \ell'} \frac{\mathcal{R}_{2 \ell', 1s}}{\Gamma_{2 \ell'}}}{\Gamma_{2 \ell}-\mathcal{R}_{2 \ell,2 \ell'} \frac{\mathcal{R}_{2 \ell', 2 \ell}}{\Gamma_{2 \ell'}}}\label{eq:C-fact}
\earr
The $C_{2\ell}$ factors generalize Peebles's $C$-factor \cite{AliHaimoud:2010dx}: they represent the effective probabilities that an atom starting in $2\ell$ reaches the ground state rather than the continuum, either directly, or after first transitioning to the other $n = 2$ state. This is best seen by rewriting them in the form
\barr
C_{2\ell} &=& \frac{\mathcal{R}_{2 \ell, 1s}+ \mathcal{R}_{2 \ell, 2 \ell'} \frac{\mathcal{R}_{2 \ell', 1s}}{\Gamma_{2 \ell'}}}{\mathcal{B}_{2 \ell} + \mathcal{R}_{2 \ell,2 \ell'} \frac{\mathcal{B}_{2 \ell'}}{\Gamma_{2 \ell'}} + \mathcal{R}_{2 \ell, 1s} +  \mathcal{R}_{2 \ell, 2 \ell'} \frac{\mathcal{R}_{2 \ell', 1s}}{\Gamma_{2 \ell'}}}. \label{eq:C2l}
\earr
These simple equations are \emph{exact}, provided that one uses exact rates $\mathcal{R}_{2 \ell, 1s}$. They form the basis of \hyrec-2, which we describe in the next Section.

\section{HyRec-2 equations}\label{sec:swift}

The computational bottleneck of the exact calculation of the recombination history comes from the evaluation of the net decay rates from the first excited states to the ground state, $\dot{x}_{2s}|_{1s}$ and $\dot{x}_{2p}|_{1s}$, or equivalently, the coefficients $\mathcal{R}_{2s, 1s}$, $\mathcal{R}_{2p, 1s}$. The basic idea of \hyrec-2~is to use a simple analytic base model for these rates, along with numerical corrections pre-tabulated with \hyrec. We now describe the simple base model.

\subsection{The base approximate model}

Neglecting stimulated 2-photon decays \cite{Chluba:2005uz}, and absorption of non-thermal photons redshifted out of the Lyman-$\alpha$ line \cite{Hirata:2008ny,Kholupenko:2006jm}, as well as Raman scattering \cite{Hirata:2008ny}, and higher-order Lyman transitions\footnote{Since the $3p$ state is very nearly in thermal equilibrium with the $2s$ state at early times, Ly-$\beta$ decays can be recast in terms of effective $2s-1s$ transitions, see \cite{AliHaimoud:2010dx}.} \cite{Chluba:2007yp,Kholupenko:2009my}, the net $2s-1s$ decay rate can be approximated as the spontaneous $2s-1s$ 2-photon decay rate $\Lambda_{2s,1s}$ \cite{Goldman:1989zz}, as was originally done in \cite{Peebles:1968ja, Zeldovich_69}:
\be
\mathcal{R}_{2s, 1s} \approx \Lambda_{2s, 1s} \approx 8.22 ~\textrm{s}^{-1}. \label{eq:2s}
\ee
In the limit of an infinitely narrow Lyman-$\alpha$ resonance, and neglecting corrections due to higher-order two-photon transitions \cite{Hirata:2008ny,Chluba:2009us,Chluba:2007qk}, frequency diffusion \cite{Hirata:2009qy, Chluba:2009gv}, and feedback from higher-order Lyman transitions \cite{Chluba:2007yp,Kholupenko:2009my}, the net $2p-1s$ decay rate can be approximately obtained with the Sobolev approximation \cite{Seager:1999km}:
\barr
\mathcal{R}_{2p, 1s} \approx R_{\rm Ly \alpha} \equiv \frac{8 \pi H}{3 \nh x_{1s} \lambda_{\rm Ly \alpha}^3}, \label{eq:2p}
\earr
where $H$ is the Hubble rate, $\lambda_{\rm Ly \alpha} \approx 1216$ \AA~ is the wavelength of the Lyman-$\alpha$ transition, and $x_{1s} \approx 1-x_e$ is the fraction of hydrogen in the ground state.

\hyrec's-\textsc{emla2s2p} mode consists in solving the 4-level equations \eqref{4level xe}-\eqref{eq:C-fact}, with $\mathcal{R}_{2s, 1s} = \Lambda_{2s, 1s}$ and $\mathcal{R}_{2p, 1s} = R_{\rm Ly \alpha}$. While this mode neglects a variety of radiative transfer effects, listed earlier, it accounts \emph{exactly} for non-equilibrium of the excited states of hydrogen, up to an arbitrarily high number of states, through the effective rates $\mathcal{A}_{2 \ell}$, $\mathcal{B}_{2 \ell}$, and $\mathcal{R}_{2\ell, 2 \ell'}$. 

\begin{figure}[ht]
\includegraphics[width=1\columnwidth]{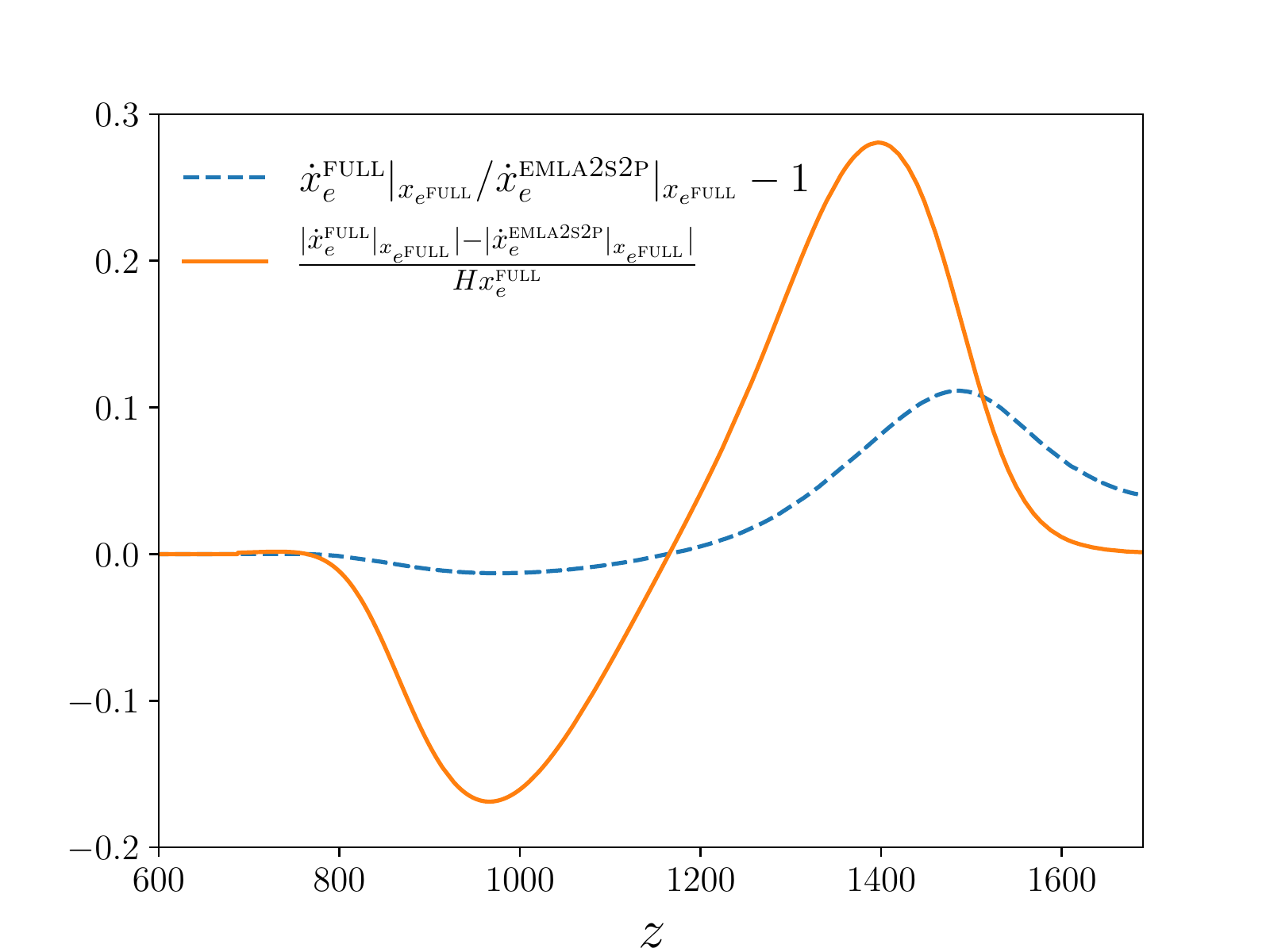}
\caption{\emph{Blue dotted curve}: Fractional difference in the rate of change of the free-electron fraction as a function of redshift between \hyrec-\textsc{emla2s2p} and \hyrec-\full. Note that this difference is computed at the same value of $x_e$. This difference shows the additional effect of solving radiative transfer equations for the photon population. \emph{Orange solid curve}: Absolute difference in the logarithmic derivative $d \ln x_e/d\ln a = \dot{x}_e/(H x_e)$. This shows that at low redshifts $z \lesssim 800$ and high redshifts $z \gtrsim 1600$ the \textsc{emla2s2p} model is accurate enough, but in the intermediate region a detailed radiative transfer calculation is important.}
\label{xe dot}
\end{figure}

Fig.~\ref{xe dot} shows the difference between the time derivatives $\dot{x}_e$ in the \full~and \textsc{emla2s2p} modes, both evaluated at the same redshift and same value of $x_e$. We see that the difference becomes negligible at $z \lesssim 800$. This is expected, as at low redshifts the net recombination rate is controlled by the efficiency of recombinations to the excited states (which are modeled exactly through the effective recombination coefficients), rather than decays to the ground state. We see that the \emph{fractional difference} $\Delta \dot{x}_e/\dot{x}_e$ (blue dotted curve) remains at the level of a few percent even at $z \sim 1700$. Nevertheless the difference $\Delta \dot{x}_e/H x_e$ (orange solid curve) becomes negligible at $z \gtrsim 1600$. As a consequence this high-redshift fractional difference does not result in significant absolute differences in the free-electron fraction, let alone observable effects in CMB anisotropies.

\subsection{Correction function}

The idea behind \hyrec-2 is simple: we want to find a correction to the net $2p-1s$ decay rate that reproduce exact calculations as accurately as possible. Our approach is similar in spirit to the analytic approximations presented in Refs.~\cite{Hirata:2008ny, Hirata:2009qy}, except the correction we compute is numerical and \emph{exact}, for a given cosmology. Similar corrections are implemented in the current version of \recfast, as well as in \textsc{recfast++} \cite{Chluba:2010ca}, but our implementation improves on both of these codes in the following ways. First and foremost, the base model of \hyrec-2~accounts exactly for the effect of highly-excited states, through the effective rates, while the base model of \recfast~and \recfast++ is Peebles' effective three-level atom. We describe this model in Appendix \ref{app:recfast} for completeness. Second, we \emph{tabulate} the corrections as a function of radiation temperature, rather than fit them with phenomenological functions as is done in \recfast. Third, we implement corrections directly at the level of the free-electron fraction \emph{derivative} $\dot{x}_e$ rather than at the level of the free-electron fraction as done in \textsc{recfast++}. Last but not least, we compute the correction function not just at a fiducial cosmology, but around it, by also tabulating its derivatives with respect to relevant cosmological parameters. 

In more detail, \hyrec-2~solves the 4-level equations \eqref{4level xe}-\eqref{eq:C-fact}, with
\be
\mathcal{R}_{2s, 1s} = \Lambda_{2s, 1s}, \ \ \ \ \ \ \mathcal{R}_{2p, 1s} = \frac{R_{\rm Ly\alpha}}{1 + \Delta(z)}. \label{eq:Delta}
\ee
The dimensionless correction $\Delta(z)$ is solved for by imposing that $\dot{x}_e^{\textsc{hyrec-2}}(z, x_e^{\full}) = \dot{x}_e^{\full}(z, x_e^{\full})$. Note that the two derivatives are evaluated at the same value of the free-electron fraction, computed in \hyrec's default \full~mode. This enforces that the two solutions are also identical (within machine precision), $x_e^{\textsc{hyrec-2}} = x_e^{\full}$. Given that the \textsc{emla2s2p} mode is obtained by setting $\Delta = 0$, the correction $\Delta$ is proportional to $\dot{x}_e^{\textsc{emla2s2p}} - \dot{x}_e^{\full}$. For completeness, we provide the explicit equation for $\Delta$ in Appendix \ref{app:Delta}.

In principle, one could define \emph{two} correction functions: one for the 2-photon decay rate $\mathcal{R}_{2s, 1s}$ in addition to the correction to the net Lyman-$\alpha$ decay rate $\mathcal{R}_{2p, 1s}$. One could solve for the two corrections by imposing that Eq.~\eqref{eq:x2l} reproduces the fractional abundances $x_{2s}$, $x_{2p}$ computed in \hyrec's \full~mode. We have opted to not follow this route, however, as one single correction function is sufficient to reproduce the exact $\dot{x}_e$. Moreover, at $z \gtrsim 800$ the populations of the excited levels depend only weakly on the rates of decay to the ground state, as they are subdominant to photoionizations and indirect transitions to the other excited state, thus the problem may be numerically ill-posed -- in other words, corrections in the $2s-1s$ and $2p-1s$ net decay rates are essentially degenerate at high redshift, thus it is more robust to only compute one single correction. 

\subsection{Cosmology dependence}

The recombination rate, thus correction function $\Delta(z)$, depend not only on redshift, but also on cosmological parameters, through the hydrogen abundance $\nh$, radiation temperature today $T_0$ and the Hubble rate $H$. It was shown in Ref.~\cite{ivanov_20} that the dependence on $T_0$ can be fully reabsorbed by expressing $H$ and $x_e$ as a function of radiation temperature $T_r = T_0 (1 +z)$, rather than redshift, and as a function of the baryon-to-photon and matter-to-photon \emph{number ratios}, proportional to the rescaled parameters
\barr
\hat{\omega}_{\rm b} &\equiv& \omega_{\rm b}(T_0^{\rm FIRAS}/T_0)^3, \\ 
\hat{\omega}_{cb} &\equiv& \omega_{cb}(T_0^{\rm FIRAS}/T_0)^3,
\earr
where $T_0^{\rm FIRAS} \equiv 2.7255$ K is the fiducial CMB temperature measured by FIRAS \cite{Fixsen:2009ug}, and $\omega_b, \omega_{cb}$ are the density parameters for baryons and baryons + cold dark matter, respectively.

The Hubble rate, expressed as a function of photon temperature, then only depends on $\hat{\omega}_{cb}$, the effective number of relativistic species $N_{\rm eff}$ (assuming the standard neutrino-to-photon temperature ratio), and neutrino masses. Given the current upper limits on the sum of neutrino masses $\sum m_\nu < 0.12$ eV \cite{Aghanim:2018eyx}, neutrinos are relativistic at the relevant redshifts $z \gtrsim 800$, thus the Hubble rate and $\Delta$ is very weakly dependent on $\sum m_\nu$. We checked explicitly that the dependence of $\Delta$ on neutrino masses is completely negligible, given the current upper limits. 

In principle, the correction function $\Delta(z)$ depends on both $\hat{\omega}_b$ and the helium mass fraction $Y_{\rm He}$: the hydrogen density is proportional to $\hat{\omega}_{\rm H} \equiv \hat{\omega}_b (1 - Y_{\rm He})$, and the evolution of the matter temperature depends on the total number density of free particles, hence on the helium-to-hydrogen number ratio $f_{\rm He} = (m_{\rm H}/m_{\rm He}) Y_{\rm He}/(1- Y_{\rm He})$. However, the matter temperature only starts departing from the radiation temperature at $z \lesssim 200$, and is in tight equilibrium with it at $z \gtrsim 800$, during which the radiative transfer correction is relevant. The dominant effect of helium abundance variations is therefore included in the parameter $\hat{\omega}_{\rm H}$, and we do not account for any dependence of $\Delta(z)$ on $Y_{\rm He}$ beyond this parameter. To be clear, the code does self-consistently include the $Y_{\rm He}$ dependence on the matter temperature evolution, but we do not propagate this dependence to $\Delta(z)$, as it is negligible at $z \gtrsim 800$. Throughout this paper $Y_{\rm He}$ is set by the BBN constraint \cite{Ade:2015xua} and not considered as a free parameter for the bias analysis in Section \ref{bias}. However, our formulation of the cosmology dependence in terms of $\hat{\omega}_H$ is fully general and allows for arbitrary values of $Y_{\rm He}$, including outside the BBN relation.

Lastly, the recombination history can be affected by a variety of processes that might have injected energy, such as particle annihilation \cite{Chluba:2009uv,Padmanabhan:2005es, Giesen_12} or decay \cite{Adams_98, Zhang:2007zzh, Mapelli:2006ej, Pierpaoli:2003rz, Chen:2003gz}, primordial black hole evaporation \cite{Poulin_17, Poulter_19} or accretion \cite{Miller_00, Ricotti_08, Ali-Haimoud_17}. These effects are accounted for in \hyrec-2~by adding source terms in the differential equations for $x_e$ and $T_m$ (see e.g.~Ref.~\cite{Giesen_12, Ali-Haimoud_17} for details). In principle, the correction function $\Delta$ also depends on these effects. For instance, $\Delta$ does depend on the dark matter annihilation parameter $p_{\rm ann} = \langle \sigma v \rangle/m_\chi$. We checked explicitly that, neglecting this dependence leads to a fractional error in $x_e$ under $3\times10^{-4}$ when $p_{\rm ann}$ is increased up to Planck's $3\sigma$ upper limit. This error is comparable to the estimated uncertainty in \hyrec~and is certainly well below the theoretical uncertainty on the effect of dark matter annihilation on the recombination history. It is therefore safe to neglect the dependence of the correction function on $p_{\rm ann}$ and other energy-injection parameters.

\begin{figure}[ht]
\includegraphics[width=1\columnwidth]{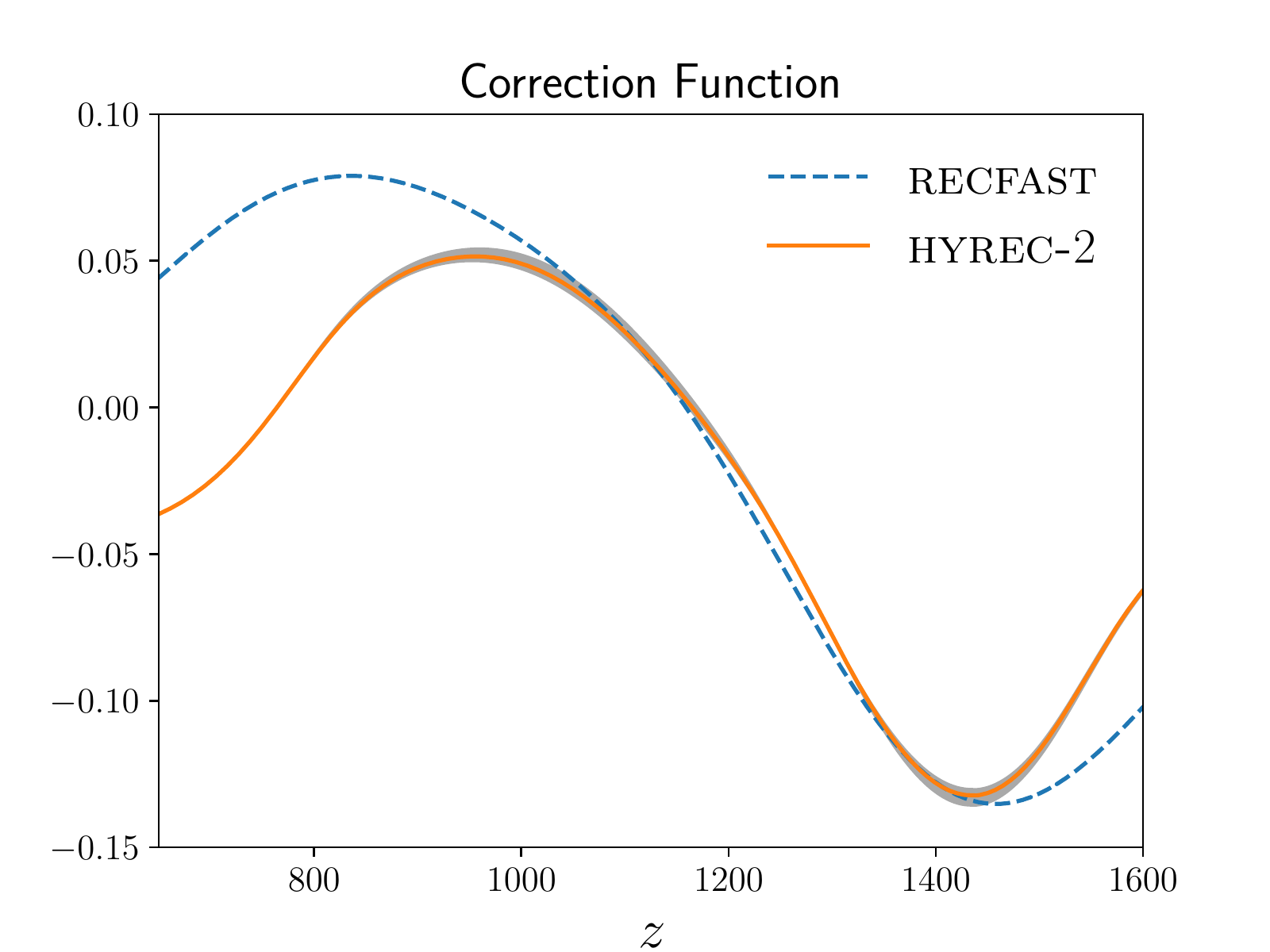}
\caption{Correction to the Lyman-$\alpha$ decay rate $\Delta(z)$ for the fiducial cosmology, defined in Eq.~\eqref{eq:Delta} (orange). Note that the correction function is technically a function of radiation temperature $\Delta(T_r)$; it is shown as a function of redsfhift for the fiducial value of $T_0 = T_0^{\rm FIRAS}$. The gray band shows the span of the correction function when varying cosmological parameters within \emph{Planck}'s full confidence region (see Fig.~\ref{fig:sample points}). For reference, the blue dashed curve shows the cosmology-independent fudge function adopted in \recfast, which is a sum of two Gaussians in redshift. Note that the base models used in \hyrec-2~and \recfast~are different, so the two correction functions dot not strictly have the same definition, see Appendix \ref{app:recfast} for more details.}
\label{fig:DKK}
\end{figure}

\begin{figure}[ht]
\includegraphics[width=1\linewidth]{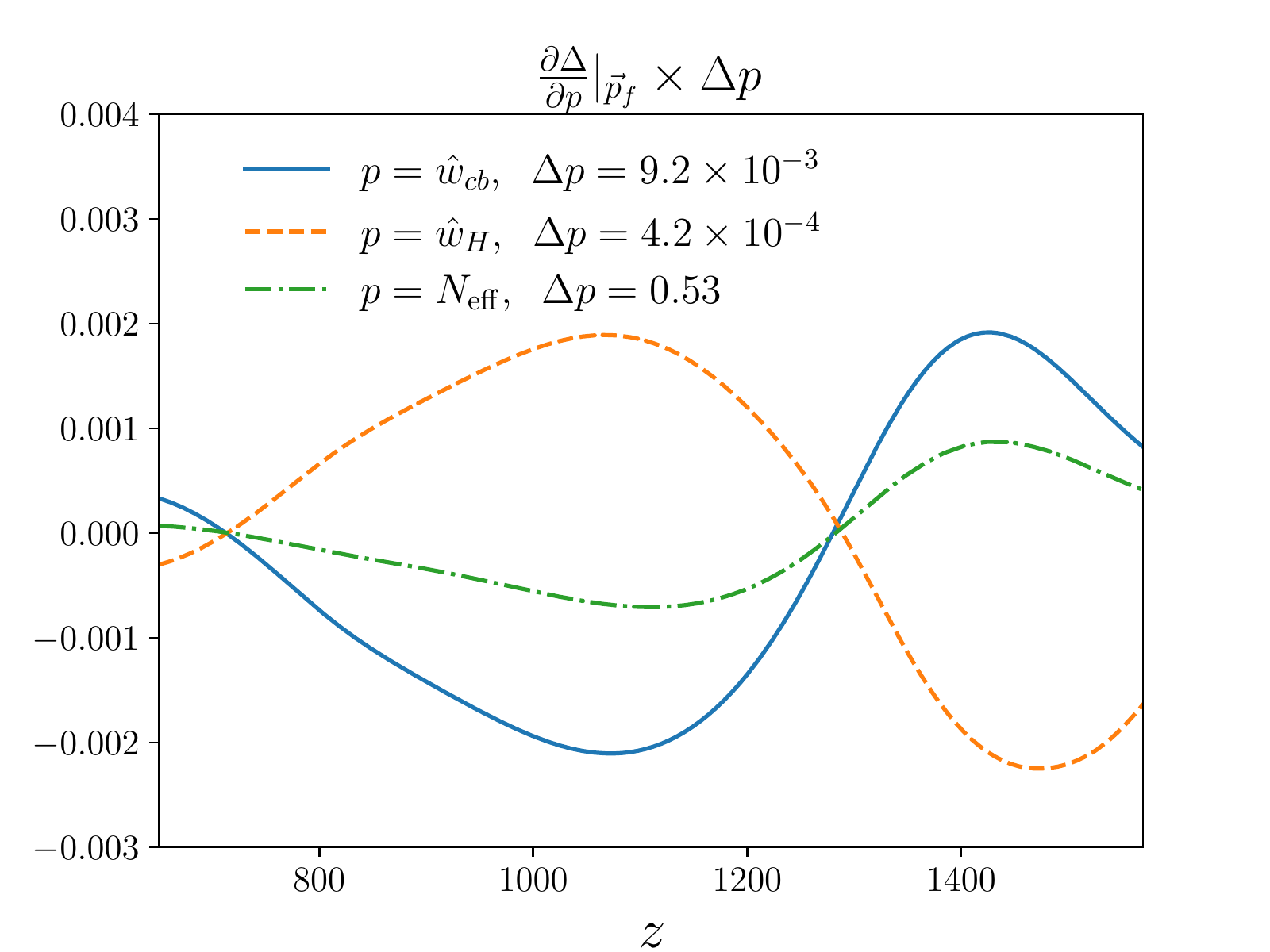}
\caption{First derivatives of the correction function $\Delta(z)$ with respect to the 3 relevant cosmological parameters $p$, multiplied by $\Delta p$ corresponding to \emph{Planck}'s $3\sigma$ confidence interval.}
\label{fig:DKK_dev}
\end{figure}

In summary, the correction function $\Delta(z) = \Delta(T_r; \vec{p})$ depends on cosmology through 3 cosmological parameters which we group in the vector $\vec{p}$:  
\be
\vec{p} \equiv (\hat{\omega}_{\rm H}, \hat{\omega}_{cb}, N_{\rm eff}).
\label{eq:p}
\ee
Since cosmological parameters are already tightly determined by CMB observations, the correction function at any set of cosmological parameters $\vec{p}$ is well approximated by a linear expansion around the \emph{Planck} best-fit parameters $\vec{p}_f$, which we refer to as the fiducial model:
\be
\Delta(T_r; \vec{p}) \approx \Delta(T_r, \vec{p}_f) + (\vec{p} - \vec{p}_f) \cdot \frac{\partial \Delta}{\partial \vec{p}}|_{\vec{p}_f}.
\ee
We therefore compute and store a total of four functions of radiation temperature, or equivalently fiducial redshift $z_f \equiv T_r/T_0^{\rm FIRAS} -1$. We list the adopted fiducial parameters in Table \ref{fiducial}. We show the function $\Delta(z, \vec{p})$ for parameters near the fiducial model in Fig.~\ref{fig:DKK}, and its derivatives with respect to cosmological parameters in Fig.~\ref{fig:DKK_dev}. We tabulate the correction functions over the fiducial redshift range $650 \leq z_f \leq 1620$.

\begin{table}[ht!]
  \centering
  \begin{tabular}{|c|c|}
  \hline
    Parameter & Fiducial Value  \\
    \hline
    $\hat{\omega}_{\rm H} \equiv \omega_b (1 - Y_{\rm He})(T_0^{\rm FIRAS}/T_0)^3$ &  0.01689\\
    $\hat{\omega}_{cb} \equiv \omega_{cb} (T_0^{\rm FIRAS}/T_0)^3$ &  0.14175\\
 $N_{\rm eff}$ & 3.046  \\
 $(m_{\nu 1}, m_{\nu 2}, m_{\nu 3})$ & (0,~0,~0.06) eV\\
    \hline
  \end{tabular}
  \caption{Cosmological parameters relevant to hydrogen recombination, along with the adopted fiducial values (derived from the Planck 2018 results \cite{Aghanim:2018eyx}), at which we compute the correction function and its first derivatives. For neutrinos, we adopt the same fiducial cosmological model as the \emph{Planck} collaboration, with two massless neutrinos, one massive neutrino with mass $0.06$ eV, and $N_{\rm eff} = 3.046$. Note that we only use derivatives of the correction function with respect to $\hat{\omega}_{\rm H}$, $\hat{\omega}_{cb}$ and $N_{\rm eff}$, as we found that its dependence on neutrino masses is negligible.}
  \label{fiducial}
\end{table}

\subsection{Numerical integrator and runtime}

The radiative transfer equation solved in \hyrec's default \full~mode is a partial differential equation, and as a consequence the timestep is tied to the frequency resolution, and must be sufficiently small to ensure convergence. For reference, the default logarithmic step in scale factor is $\Delta \ln a = 8.49 \times 10^{-5}$ (to compute the correction functions at high redshift, we used an even smaller timestep for increased accuracy). On the other hand, \hyrec-2~only solves an ordinary differential equation (ODE), and the timestep can be considerably increased at no noticeable cost in accuracy, provided one uses a sufficiently high-order numerical integrator. We found that we could safely increase the logarithmic step in scale factor to $\Delta \ln a = 4 \times 10^{-3}$ at virtually no loss of accuracy, using a 3rd order explicit integrator. At early times, when the ODE is stiff, we use an expansion around the Saha equilibrium solution (see Ref.~\cite{AliHaimoud:2010dx}). To make the code stable we use a smaller time step during and slightly after this phase. With our setup, we checked that the fractional difference in $x_e$ due to the increased timestep is less than $10^{-4}$ at all redshifts, comparable to the estimated uncertainty in \hyrec. 

The simple ODE solved in \hyrec-2, combined with a larger timestep, considerably reduces the recurring computation time, to less than 1 millisecond per cosmological model on a standard laptop, see Tab.~\ref{tab:running time} for a comparison with \hyrec-\full~and \recfast. With this short run time, the recombination history calculation is never the bottleneck of CMB anisotropy Boltzmann codes.

\begin{table}[ht!]
  \centering
  \begin{tabular}{|c|c|c|c|}
  \hline
    Code & ~~~\hyrec ~~~& ~~~\hyrec-2 ~~~ & ~~~\recfast~~~\\
    \hline
    Run time (ms) & 409 & 0.76 & 23\\
  \hline
  \end{tabular}
  \caption{Default run time of each code. Note that this is the \emph{recurring} run time, which does not account for the loading of data in \hyrec-2, as this needs to be done once and for all. The run times are calculated on a standard laptop (2.0 GHz Intel i5 processor, 16 GB of RAM).} 
\label{tab:running time}
\end{table}

\section{Accuracy of \hyrec-2} \label{sec:results}

\subsection{Sample cosmologies}

\begin{figure*}[ht]
\includegraphics[width=2\columnwidth]{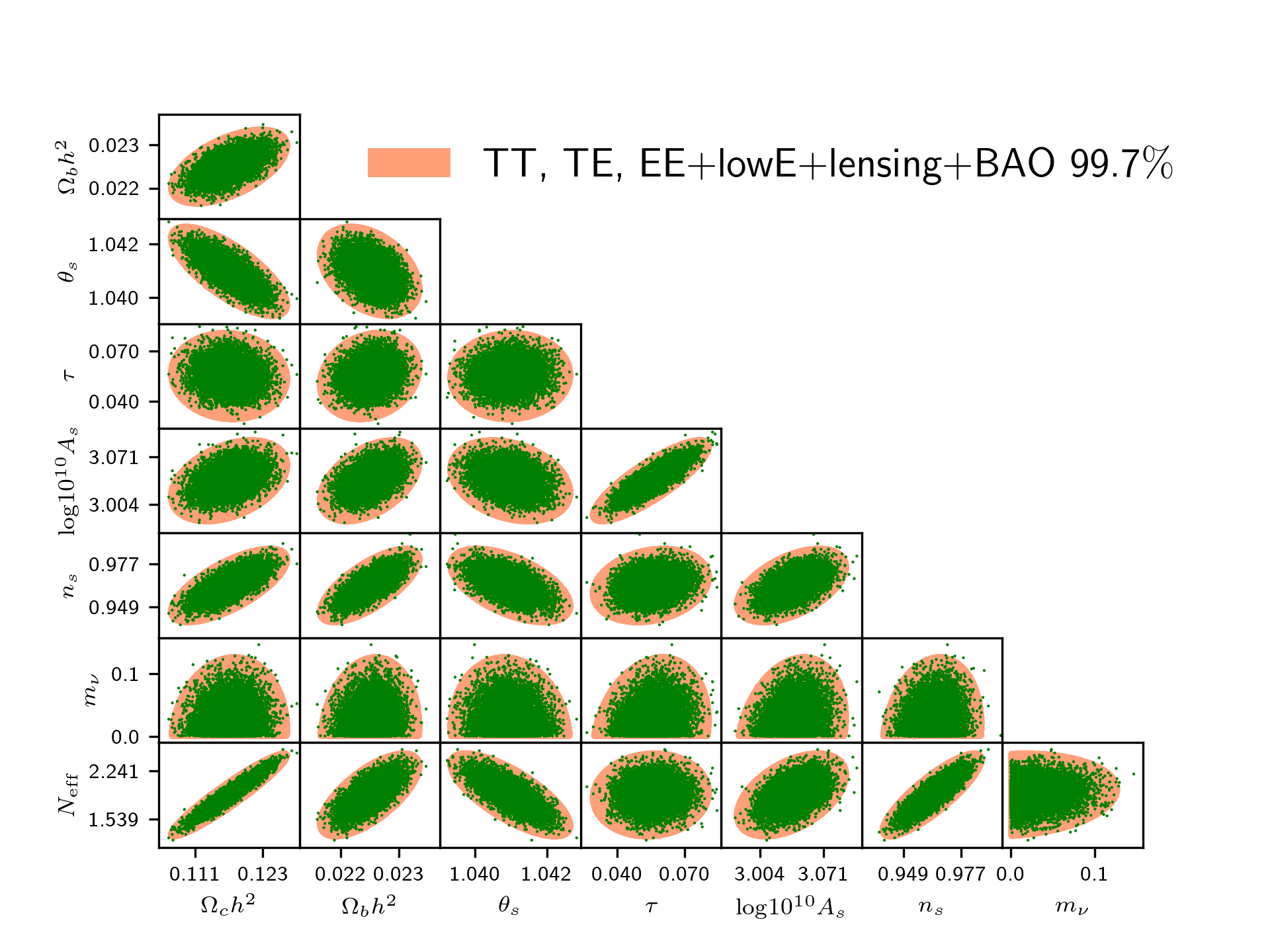}
\caption{Ten thousands sample cosmologies used to check the accuracy of \hyrec-2, and for the bias analysis in Fig.~\ref{fig:bestfit}. These samples are drawn from a Gaussian distribution with covariance matrix provided by the \emph{Planck} collaboration \cite{Aghanim:2018eyx}. As expected most of samples are within 99.7\% confidence region.}
\label{fig:sample points}
\end{figure*}

To check the accuracy of \hyrec-2, we randomly generated ten thousand sample cosmologies from the 8-dimensional Gaussian likelihood derived from the \emph{Planck} 2018 covariance matrix \cite{Aghanim:2018eyx} (TT, TE, EE+lowE+lensing+BAO, 2-parameter extension)\footnote{base\_nnu\_mnu\_plikHM\_TTTEEE\_lowl\_lowE\_lensing\_BAO.covmat from https://wiki.cosmos.esa.int/planck-legacy-archive/index.php/Cosmological\_Parameters.}, shown in Fig.~\ref{fig:sample points}. As can be seen in Fig.~\ref{fig:sample points}, most samples are within the 99.7\% confidence region, and there are a handful of samples outside, as expected. We use these sample cosmologies to check the accuracy of \hyrec-2 compared to the reference model, the original \hyrec.

\subsection{Accuracy of the free-electron fraction}

\begin{figure}[ht]
\includegraphics[width=0.9\columnwidth]{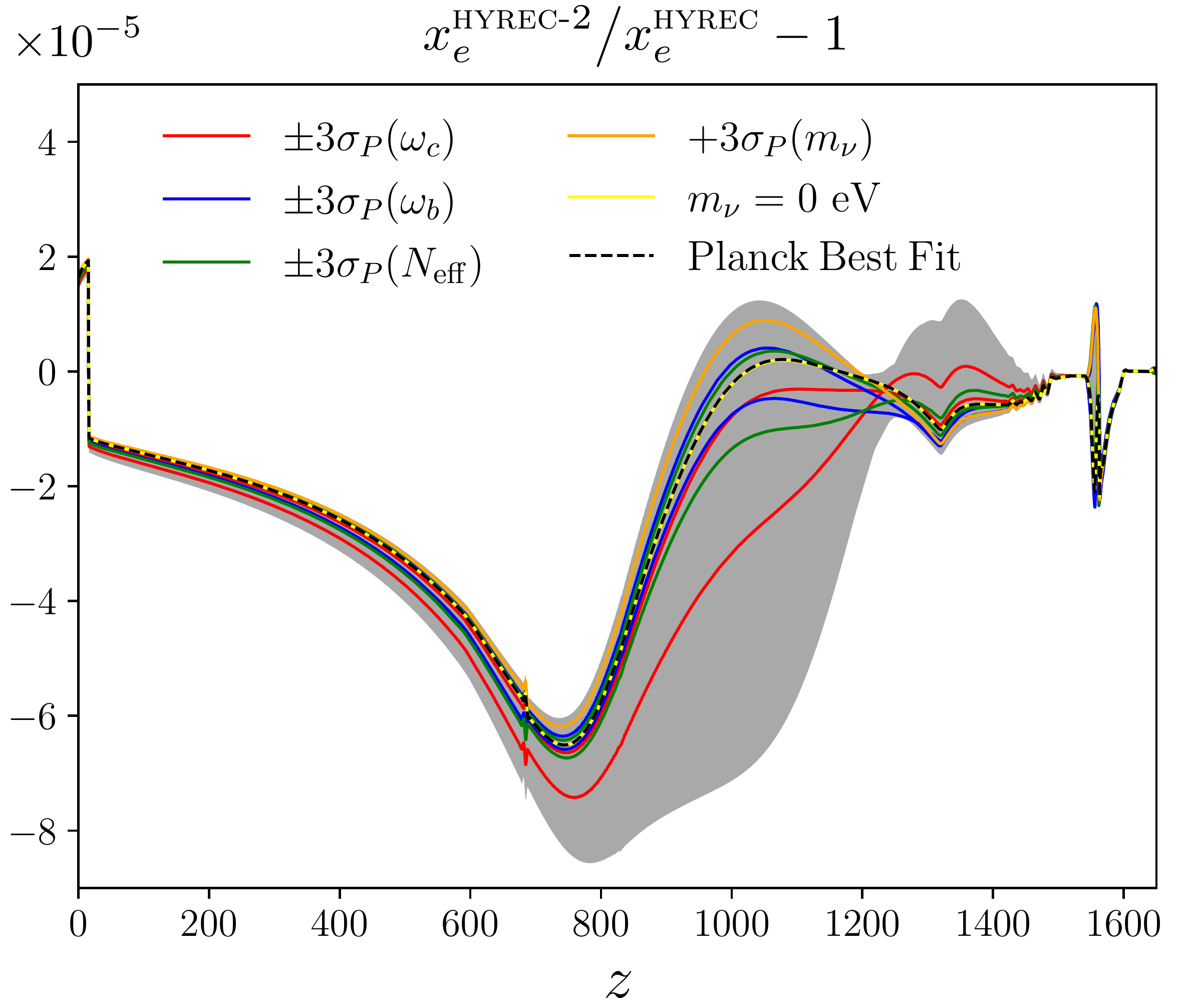}
\includegraphics[width=0.95\columnwidth]{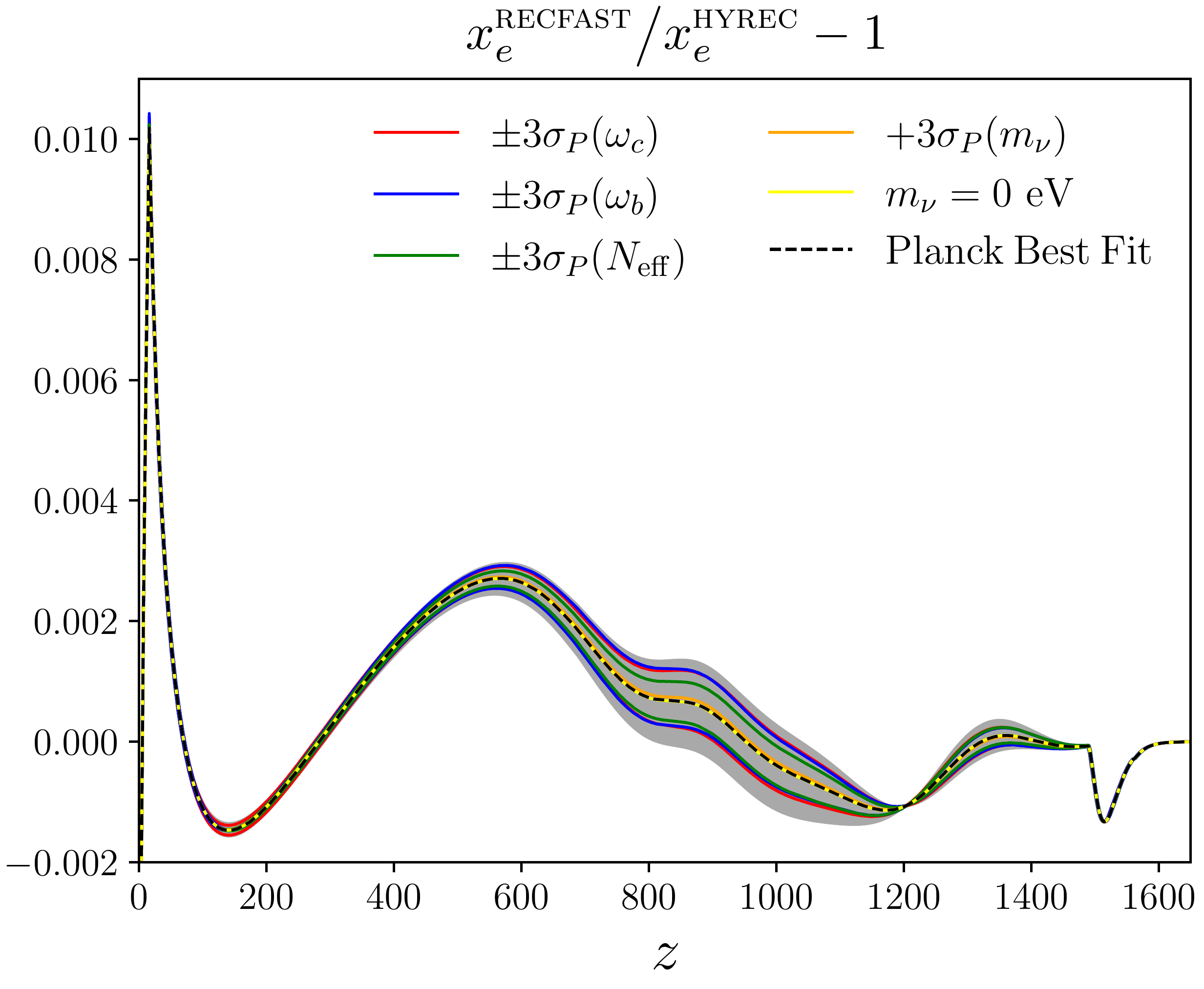}
\caption{Fractional differences in the free-electron fraction $x_e$ of \hyrec-2 (upper panel) and \recfast~ (bottom panel) with respect to the reference model \hyrec-\full. The small ($< 10^{-4}$) fractional difference between \hyrec-2 and \hyrec~for the Planck best-fit cosmology is due to the different timesteps in the two codes (moreover the correction function were computed using a higher-accuracy mode of \hyrec, with a smaller-than-default timestep). The differences are calculated by changing each parameter with \emph{Planck} $\pm 3\sigma$. The shaded area corresponds to the differences calculated with the 10,000 sample cosmologies shown in Fig.~\ref{fig:sample points}. }
\label{fig:xe}
\end{figure}

The fractional difference in the free-electron fraction $x_e(z)$ computed in \hyrec~and \hyrec-2 is shown in the top panel of Fig.~\ref{fig:xe}, for a broad range of cosmological parameters. 
As the plots show, the fractional difference is less than $10^{-4}$ when cosmological parameters are varied within \emph{Planck}'s 99.7\% confidence region. This is lower than the estimated uncertainty in \hyrec. 

The $\sim 5\times 10^{-5}$ feature at $z \approx 1600$ is due to the different times at which stiff approximations are turned on, and has no observational consequence whatsoever. Note that even though we neglect the dependence of the correction function on neutrino masses, the code remains accurate even when they are varied away from their fiducial values, within Planck's $3 \sigma$ limits.  

For comparison, we show the fractional difference between \recfast~and \hyrec~in the lower panel of Fig.~\ref{fig:xe}, for the same parameters. This difference is up to two orders of magnitude larger: it gets as large as $\sim 4 \times 10^{-3}$ at $z \gtrsim 200$, and grows to $\sim 1\%$ at $z \lesssim 100$. As we will show in more detail in Section \ref{bias}, this difference is negligible for \emph{Planck}, but can lead to non-trivial biases for next-generation CMB experiments. 

\subsection{Bias of cosmological parameters} \label{bias}

\begin{figure}[ht]
\includegraphics[width=0.95\columnwidth]{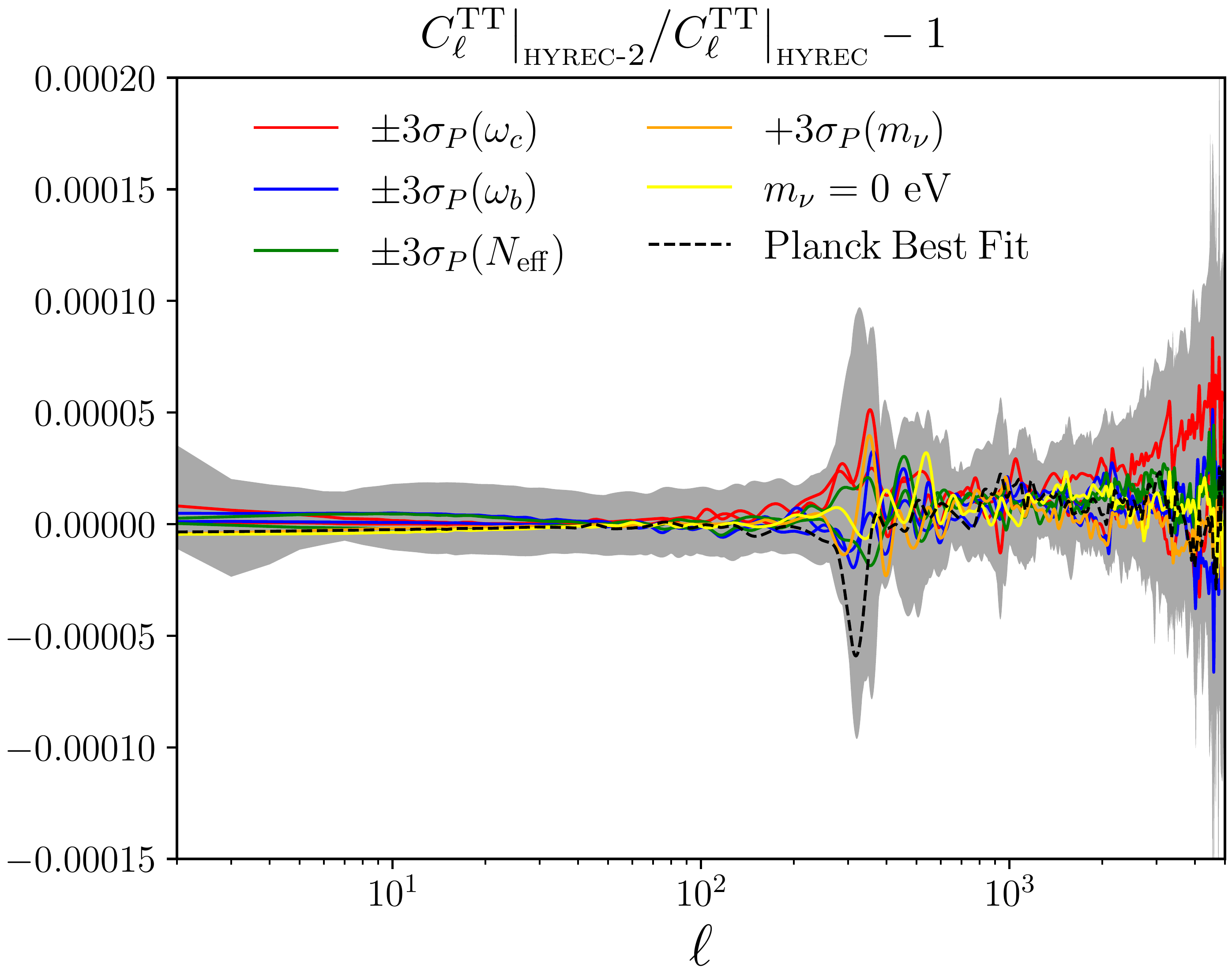}
\includegraphics[width=0.95\columnwidth]{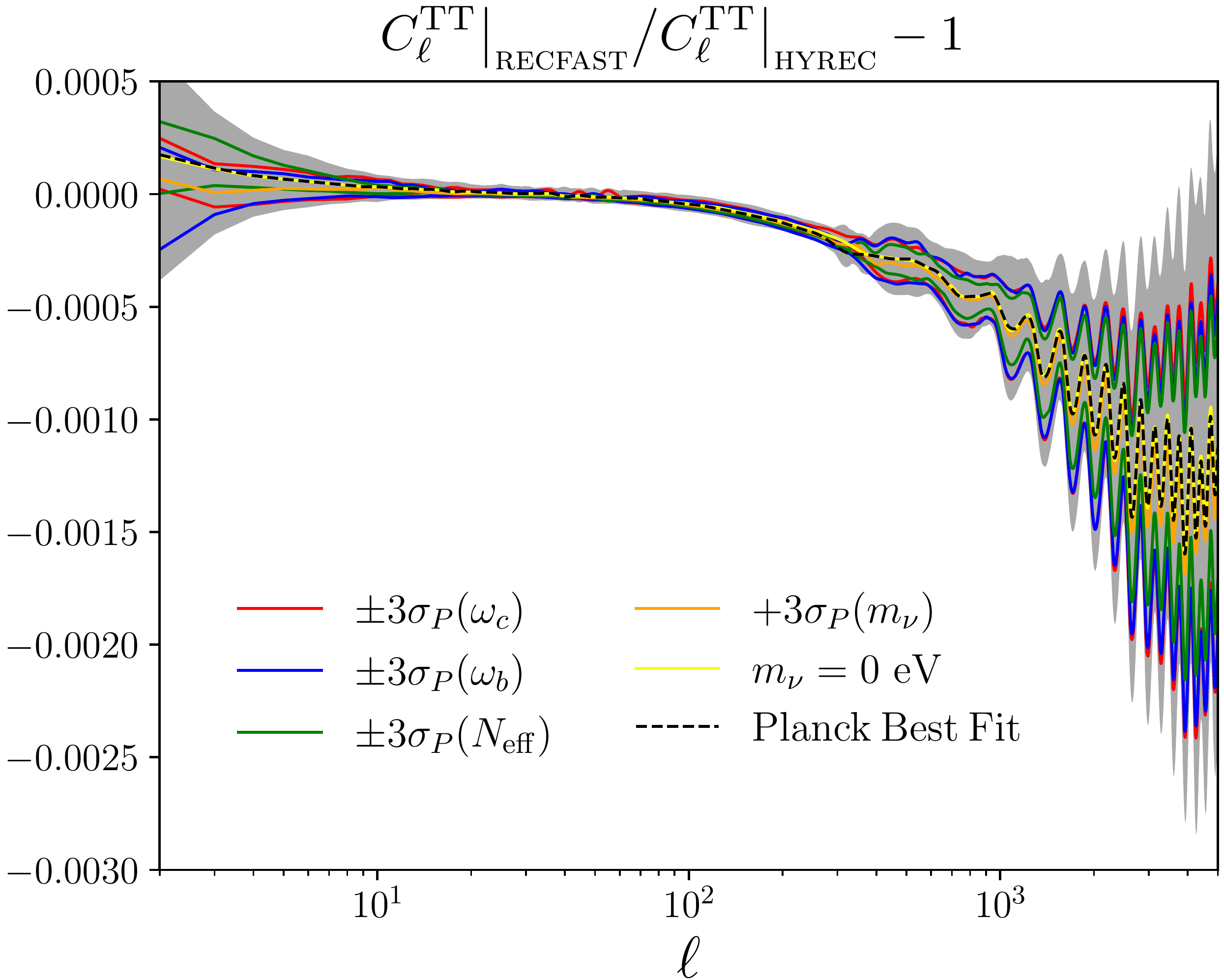}
\caption{Fractional error in $C_\ell^{TT}$ when using \hyrec-2 (top) or \recfast~(bottom) instead of \hyrec~(in its default \full~mode). The shaded area corresponds to the differences calculated with the 10,000 sample cosmologies shown in Fig.~\ref{fig:sample points}. In all cases, the $C_{\ell}$'s are computed with the Boltzmann code \textsc{class} \cite{Lesgourgues:2011re}. Note that with the default precision settings of \textsc{class}, for a few sample cosmologies the $C_{\ell}$'s showed relatively large errors at low-$\ell$ (still $\lesssim 4\times 10^{-4}$); we checked that those errors disappear once the precision of \textsc{class} is increased.}
\label{fig:Cl-diff}
\end{figure}

The metric with which the accuracy of an approximate recombination code is to be measured is the \emph{biases} it induces on cosmological parameters. In the limit of small errors, these biases are directly proportional to the error in CMB anisotropy angular power spectra, $C_{\ell}$. For illustration, we show in Fig.~\ref{fig:Cl-diff} the error in the temperature power spectrum $C_{\ell}^{\rm TT}$ for a variety of cosmological parameters varied within the \emph{Planck} 99.7\% confidence region. We see that \hyrec-2 is more accurate than \recfast~by more than one order of magnitude at small angular scales.

We can estimate the biases from a simple Fisher analysis. We denote by $\bs{C} \equiv \{ C_{\ell}^{\rm TT}, C_{\ell}^{\rm TE}, C_{\ell}^{\rm EE}, C_{\ell}^{\rm dd} \} $ the vector containing all the temperature and polarization power spectra and cross-spectra, as well as the power spectrum of lensing deflection. We denote by $\bs{\Sigma}$ their covariance matrix, which we describe in more detail below. The chi-squared of a set of cosmological parameters $\vec{p}$ is
\be
\chi^2(\vec{p}) = \left(\bs{C}(\vec{p}) - \hat{\bs{C}} \right) \cdot \bs{\Sigma}^{-1} \cdot \left(\bs{C}(\vec{p}) - \hat{\bs{C}} \right), 
\ee
where $\hat{\bs{C}}$ is an estimator of $\bs{C}$ constructed from the data, with covariance $\bs{\Sigma}$. The best-fit cosmology $\vec{p}_{\rm bf}$ is found by minimizing the $\chi^2$. Taylor-expanding around some fiducial cosmology $\vec{p}_0$, and neglecting the terms proportional to second derivatives of $C_{\ell}$ \cite{Dodelson:2003ft} we get
\barr
p_{\rm bf}^i - p_0^i &=& \bs{B}^i(\vec{p}_0) \cdot \left(\bs{C}(\vec{p}_0) - \hat{\bs{C}} \right),\\
\bs{B}^i &\equiv& - (F^{-1})^{ij} \frac{\partial \bs{C}}{\partial p^j} \cdot \bs{\Sigma}^{-1}. \label{eq:B}
\earr
where $F_{ij}$ is the Fisher matrix, whose inverse is the covariance of the best-fit cosmological parameters, and whose elements are 
\be
F_{ij} = \frac{\partial \bs{C}}{\partial p^i}\cdot \bs{\Sigma}^{-1} \cdot \frac{\partial \bs{C}}{\partial p^j}.
\ee
Suppose the data is a (noisy) realization of the cosmology $\vec{p}_0$, i.e.~that, upon averaging over realizations, $\langle \hat{\bs{C}} \rangle = \bs{C}(\vec{p}_0)$. If the theoretical model is unbiased, then the best-fit parameters are also unbiased, i.e.~such that, on average over realizations, $\langle \vec{p}_{\rm bf} - \vec{p}_0 \rangle = 0$. 

Now suppose that the theoretical model for $\bs{C}(\vec{p})$ has a systematic error $\Delta \bs{C}$:
\be
\bs{C}(\vec{p}) = \bs{C}_{\rm true}(\vec{p}) + \Delta \bs{C}(\vec{p}).
\ee
The biased theoretical model leads to a \emph{systematic bias} in the best fit, with average
\barr
\langle p_{\rm bf}^i - p_0^i \rangle = \bs{B}^i(\vec{p}_0) \cdot \Delta \bs{C}(\vec{p}_0)\\
\approx \bs{B}^i(\vec{p}_{\rm fid}) \cdot \Delta \bs{C}(\vec{p}_0),
\label{eq:bias}
\earr
where $\bs{B}^i$ was given in Eq.~\eqref{eq:B}. For simplicity we approximated $\bs{B}(\vec{p}_0) \approx \bs{B}(\vec{p}_{\rm fid})$ in Eq.\eqref{eq:bias}; this will not affect the results since $\Delta \bs{C}(\vec{p}_0)$ is already a small quantity. The error to this approximation would be a small correction to a correction. The advantage of this approximation is that we need to compute $\bs{B}$ only at the fiducial cosmology.

Let us now evaluate these systematic biases for a few idealized CMB observations. The covariance matrix $\bs{\Sigma}$ has components \cite{BenoitLevy:2012va}
\barr
\Sigma_{\ell \ell'}^{XY, WZ} &\equiv& \textrm{cov}[\hat{C}_{\ell}^{\rm XY}, \hat{C}_{\ell'}^{\rm WZ}] \nonumber\\
&=& \delta_{\ell \ell'} \frac{\tilde{C}_{\ell}^{XW}\tilde{C}_{\ell}^{YZ} + \tilde{C}_{\ell}^{XZ}\tilde{C}_{\ell}^{YW}}{f_\text{sky}(2l+1)},
\earr
where, for $X = T, E, d$, 
\be
\tilde{C}^{XW}_{\ell} \equiv C_{\ell}^{XW} + \delta_{XW} N_{\ell}^{XX}, 
\ee
where $N_{\ell}^{\rm XX}$ is the instrumental noise, of the form \cite{Abazajian:2016yjj}
\be
N_{\ell}^{\rm XX} = N_0^{\rm XX} \exp \left(\frac{\ell (\ell +1) \theta_{\rm X}^2}{8 \ln 2}\right).
\ee
We adopt the noise parameters of Ref.~\cite{Shaw:2011ez} for \emph{Planck} and of Ref.~\cite{Green:2016cjr} for a CMB stage-IV experiment, which we summarize in Tab.~\ref{tab:noise}. Further, we consider an idealized Cosmic Variance Limited (CVL) case for which we assume no instrumental noise in both temperature and polarization up to $\ell = 5000$ and full sky $f_{\rm sky}=1$. In both cases the lensing reconstruction noises are calculated using the code developed by \cite{Peloton:2016kbw}.

\begin{table}[ht!]
  \centering
  \begin{tabular}{|c|c|c|c|}
  \hline
    Experiment &  Stage-IV  & CVL \\
    \hline
    $N_0^{\rm TT} (\mu\rm{K}^2$) &  $3.38\times10^{-7}$ & 0 \\ 
    $N_0^{\rm EE} (\mu\rm{K}^2$) &  $6.77\times10^{-7}$ & 0 \\
    $\theta_T, \theta_E$ (arcmin) &  1& $\ell_{\max} = 5000$\\
    $f_{\rm{sky}}$ & 0.4 & 1\\ 
  \hline
  \end{tabular}
  \caption{Noise parameters and the fraction of sky adopted in the Fisher matrix estimates.}
\label{tab:noise}
\end{table}

We randomly choose various cosmologies from \emph{Planck} full confidence level as shown in Fig.~\ref{fig:sample points} and fit each input data (\hyrec-\full~mode) to get the best fit of cosmological parameters using each code, \recfast~and \hyrec-2. We first checked that if we use the \emph{Planck} settings \cite{Aghanim:2018eyx}, both codes lead to biases well below statistical uncertainties, which confirms that \recfast~is good enough for \emph{Planck} data as established in \cite{Ade:2015xua,Aghanim:2018eyx}.

\begin{figure*}[ht]
\includegraphics[width=1.7\columnwidth]{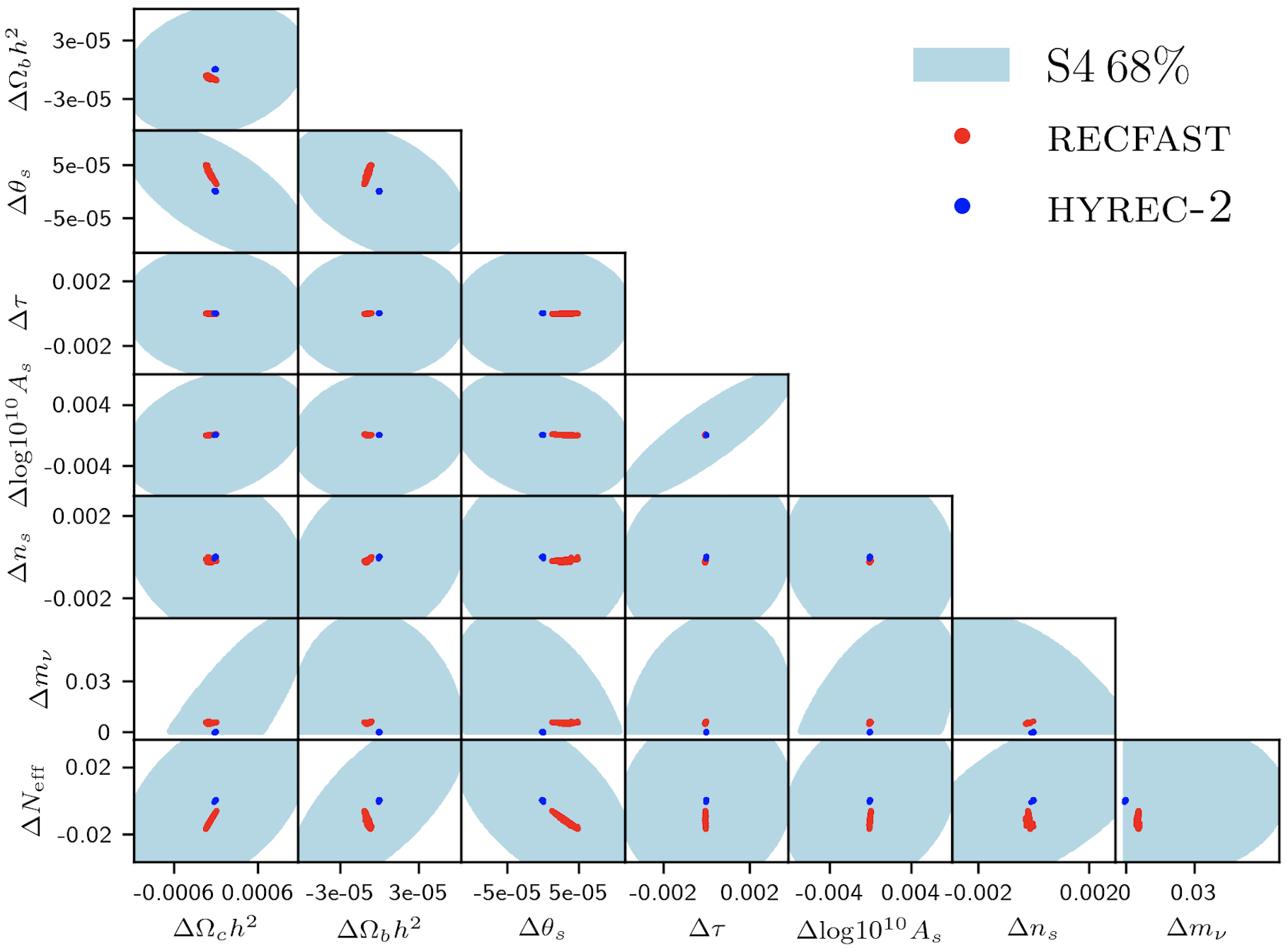}
\includegraphics[width=1.7\columnwidth]{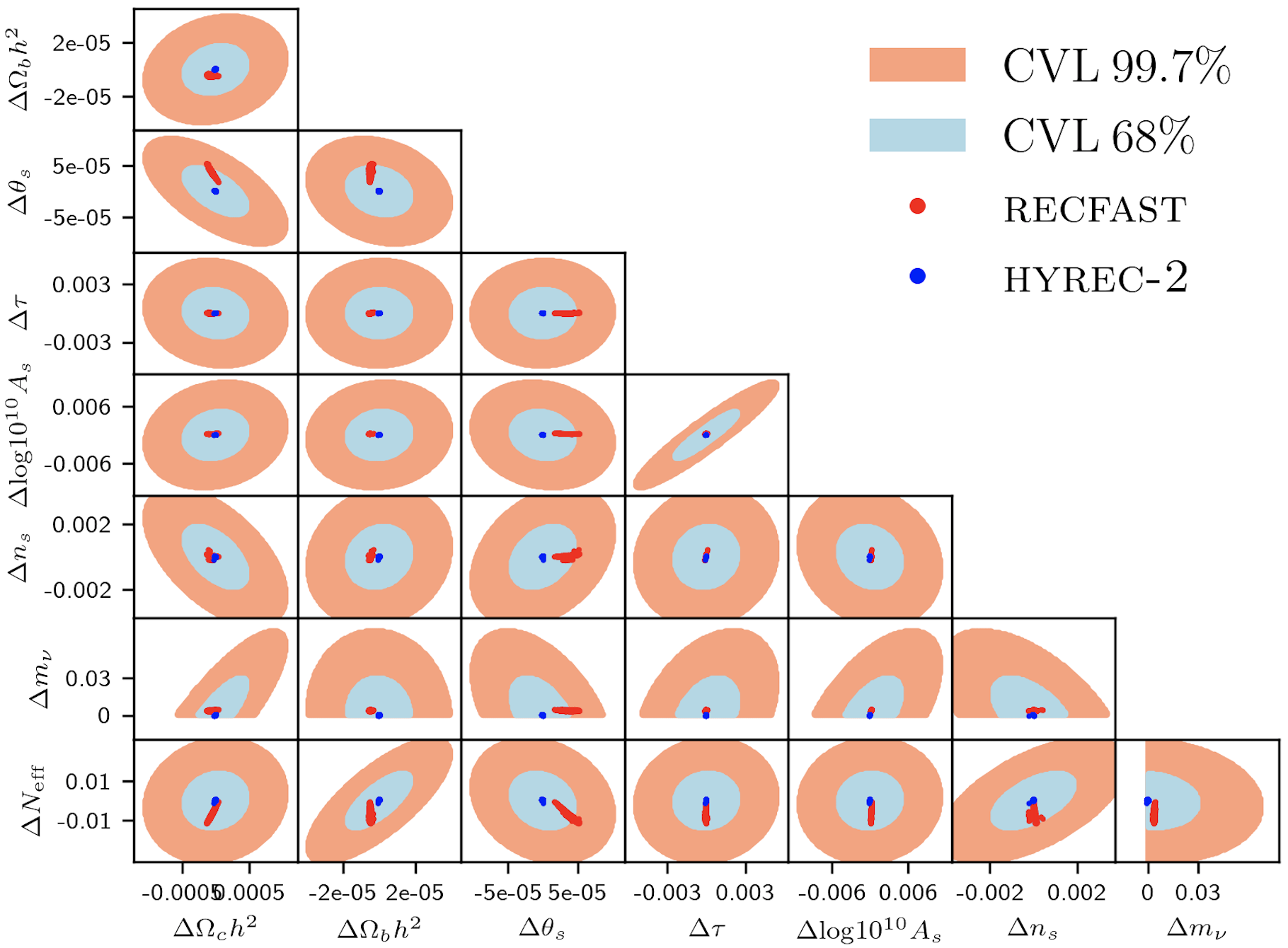}
\caption{Bias in cosmological parameters when using \hyrec-2 and \recfast, assuming \hyrec~(in its default \textsc{full} mode) provides the exact model. The top panel corresponds to CMB Stage-4 settings, and the bottom panel to an idealized CMB experiment, cosmic-variance limited (CVL) in temperature and polarization up to $\ell_{\max} = 5000$. In both cases, \hyrec-2 generates biases that are negligibly small relative to the statistical uncertainty. For a CMB Stage-4 setup, the biases from \recfast~remain within the 68\% confidence region. For the CVL experiment, however, \recfast~leads to biases reaching beyond the 68\% confidence region in some instances.}
\label{fig:bestfit}
\end{figure*}

The difference between the best fit parameters and the input parameters are shown in Fig.~\ref{fig:bestfit}. The results in the upper panel of Fig.~\ref{fig:bestfit} are obtained with the CMB S-4 setting described in Ref.~\cite{Abazajian:2016yjj}, which is $2 \leq \ell \leq 3000$ for $TT$ and $2 \leq \ell \leq 5000$ for $TE$, $EE$, and $dd$. The bottom panel shows the corresponding biases for the idealized experiment, assumed to be CVL for $2 \leq \ell \leq 5000$ both in intensity and polarization. Note that in principle the Gaussian approximation for the $C_{\ell}'s$ (on which the simple $\chi^2$ analysis implicitly relies) is inaccurate at low $\ell$; however, we checked that by changing $\ell_{\rm min}$, i.e. $\ell_{\rm min}=10$, the low multipoles do not contribute much to the biases and should not greatly affect the answer. We see from Fig.~\ref{fig:bestfit} that in some cases, biases fall outside the 68\% confidence region of the CVL experiment when using \recfast. With \hyrec-2, all biases remain much smaller than statistical uncertainties, for the full range of cosmologies allowed within the 99.7\% confidence region of \emph{Planck}.

\section{Conclusion}\label{sec:conclusion}
We have developed the new recombination code \hyrec-2, which combines high accuracy with extreme computational efficiency. This new code is as accurate as the original \hyrec~across the full range of currently allowed cosmological parameters, and is 30 times faster than \recfast, with a recurring runtime under one millisecond on a standard laptop. This makes \hyrec-2 the fastest recombination currently available, by far.

\hyrec-2 is based on an effective 4-level atom model, which captures exactly the late-time $(z \lesssim 800)$ recombination dynamics. Radiative transfer effects, which are relevant at early times, are accounted for through a numerical correction to the Lyman-$\alpha$ net decay rate, tabulated as a function of temperature using \hyrec. In order to achieve sub-0.01\% accuracy across a broad range of cosmologies, we also tabulated the derivatives of the correction function with respect to cosmological parameters. We have checked explicitly that the $\sim 10^{-4}$ fractional differences with \hyrec~result in no bias for any cosmological parameters for current, planned, and even futuristic CMB missions, for which \recfast~would not be sufficiently accurate. 

Our new recombination code will be most useful for fast and accurate CMB-anisotropy calculations, required to extract unbiased cosmological parameters from CMB-anisotropy data. In addition, it will be a key tool to study the CMB signatures of dark matter decay or annihilation \cite{Chluba:2009uv,Padmanabhan:2005es, Giesen_12} other sources of energy injection \cite{Adams_98, Chen:2003gz, Ali-Haimoud_17}, or in general any non-standard physics that may affect the recombination and thermal history \cite{Ali-Haimoud_11_metals}. 

Last but not least, in combination with the effective conductance method \cite{Ali-Haimoud_13}, \hyrec-2 can be used to efficiently compute the cosmological recombination spectrum \cite{Chluba_16}. This minute but rich signal is a guaranteed distortion to the CMB blackbody spectrum \cite{Rubino-Martin_06, Sunyaev_08}. Looking ahead, it may eventually become a powerful probe of the early Universe \cite{Chluba_19, Chluba_19b, Delabrouille_19, Sarkar_20}, complementing CMB anisotropies and opening up a new window into the Universe's early thermal history.

\section*{Acknowledgements}

We thank Jens Chluba, Antony Lewis and Julien Lesgourgues for useful conversations. This work is supported by NSF Grant No. 1820861 and NASA Grant No. 80NSSC20K0532.

\begin{appendix}

\section{Explicit expression for the correction function}\label{app:Delta}

Equation \eqref{4level xe} can be rewritten as $\dot{x}_e = - \sum_{\ell} C_{2 \ell} X_{2 \ell}$, where 
\be
X_{2 \ell} \equiv \nh x_e^2 \mathcal{A}_{2 \ell} - g_{2 \ell} x_{1s} \rme^{- E_{21}/T_r} \mathcal{B}_{2 \ell}.
\ee
In \hyrec-2, $\mathcal{R}_{2s, 1s} = \Lambda_{2s, 1s}$ and $\mathcal{R}_{2p, 1s} = R_{\rm Ly \alpha}/(1 + \Delta)$; the \textsc{emla2s2p} mode has $\Delta = 0$. We therefore have
\be
\dot{x}_e^{\hyrec-2}(\Delta) = \frac{\dot{x}_e^{\textsc{emla}} + A \Delta}{1 + B \Delta},
\ee
with 
\barr
A &=& -\frac{\Lambda_{2s,1s}\left[(\mathcal{B}_{2p} + \mathcal{R}_{2p,2s})X_{2s} + \mathcal{R}_{2p,2s}X_{2p}\right]}{\Gamma_{2s}\Gamma_{2p}-\mathcal{R}_{2s,2p}\mathcal{R}_{2p,2s}}, \\
B &=& \frac{\Gamma_{2s}(\mathcal{B}_{2p} + \mathcal{R}_{2p,2s}) -\mathcal{R}_{2s,2p}\mathcal{R}_{2p,2s}}{\Gamma_{2s}\Gamma_{2p}-\mathcal{R}_{2s,2p}\mathcal{R}_{2p,2s}}.
\earr
The correction $\Delta$ is set such that $\dot{x}_e^{\hyrec-2}(\Delta) = \dot{x}_e^{\textsc{full}}$. Solving, we find
\be
\Delta = \frac{\dot{x}_e^{\textsc{emla}} - \dot{x}_e^{\textsc{full}}}{B \dot{x}_e^{\textsc{full}} - A}.
\ee

\section{Equations for \textsc{RECFAST} in our notation} \label{app:recfast}

\subsection{Peebles' effective three-level model}

Peebles' effective 3-level model \cite{Peebles:1968ja} relies on two additional assumptions relative to the effective 4-level model that we use. First, the two states $2s, 2p$ are assumed to be in thermal equilibrium, $x_{2s} = x_{2p}/3 \equiv x_2/4$, with $x_2 \equiv x_{2s} + x_{2p}$. The recombination rate \eqref{eq:xe-dot-exact} then simplifies to 
\barr
\dot{x}_e &=& x_2 \mathcal{B}_B - n_H x_e^2 \mathcal{A}_B, \label{eq:xedot-x2}\\
\mathcal{B}_B &\equiv& \frac14 \left(\mathcal{B}_{2s} + 3 \mathcal{B}_{2p}\right),\\ 
\mathcal{A}_{\rm B} &\equiv& \mathcal{A}_{2s} + \mathcal{A}_{2p}.
\earr
The population $x_2$ of the first excited state is then obtained by solving the steady-state equation,
\barr
0 \approx \dot{x}_2 &=& n_H x_e^2 \mathcal{A}_B - x_2 \mathcal{B}_B + \dot{x}_{2s}|_{1s} + \dot{x}_{2p}|_{1s} \nonumber\\
&\approx& n_H x_e^2 \mathcal{A}_B - x_2 \mathcal{B}_B \nonumber\\
&+& \frac14\left(\Lambda_{2s, 1s} + 3 R_{\rm Ly \alpha} \right) \times \left(4 x_{1s} \rme^{- E_{21}/T_r} - x_2\right),~~~~
\earr
where here again we used the simple approximations \eqref{eq:2s} and \eqref{eq:2p} for the net decay rates to the ground state. This equation can be easily solved for $x_2$, which, upon insertion into Eq.~\eqref{eq:xedot-x2}, gives the closed form
\be
\dot{x}_e = -C \left(n_H x_e^2 \mathcal{A}_B -4x_{1s}\mathcal{B}_B e^{-E_{21}/T_r} \right),
\ee
where the Peebles $C$ factor is given by 
\be
C\equiv \frac{\Lambda_{2s,1s} + 3 R_{\text{Ly}\alpha}}{4 \mathcal{B}_B + \Lambda_{2s,1s} + 3 R_{\text{Ly}\alpha}}.
\ee
This result can also be obtained from Eqs.~\eqref{4level xe} and \eqref{eq:C2l} using the detailed balance relation $\mathcal{R}_{2s, 2p} = 3 \mathcal{R}_{2p, 2s}$, and assuming that $\mathcal{R}_{2s, 2p} \gg \Lambda_{2s, 1s}+ \mathcal{B}_{2s}$ and $\mathcal{R}_{2p, 2s} \gg R_{\rm Ly \alpha}+ \mathcal{B}_{2p}$. This assumption is required to enforce equilibrium between $2s$ and $2p$ regardless of the relative values of the other rates, and implies $C_{2s} = C_{2p} = C$. In practice, it does not hold at low enough temperature, $z\lesssim 700$. 

In addition to this equilibrium assumption, the effective rates $\mathcal{A}_B(T_m, T_r)$ and $\mathcal{B}(T_r)$ are approximated as follows:
\barr
\mathcal{A}_B(T_m, T_r) &\approx& \alpha_B(T_m) \equiv \mathcal{A}_B(T_m, 0),\\
\mathcal{B}_B(T_r) &\approx& \beta_B(T_r) \equiv \frac{(2\pi \mu_e T_r)^{3/2}}{4 h^3}e^{E_2/T_r} \alpha_B(T_r),~~~~
\earr
where $E_2 \approx - 3.4$ eV is the energy of the first excited state. In other words, the effective recombination coefficient is computed in the zero-radiation-temperature limit and the photoionization rate is assumed to be given by detailed balance, even though this is not self-consistent with the zero-radiation-temperature assumption.

\subsection{Fudge factors and functions}

Since the zero-radiation-temperature effective recombination coefficient systematically under-estimates the exact effective recombination coefficient, the code \recfast~ introduces a ``fudge factor" $F > 1$, and substitutes $\alpha_B \rightarrow F \times \alpha_B$. The fudge factor was first estimated to $F \approx 1.14$ when enforcing equilibrium between angular momentum substates of excited states \cite{Seager:1999km, Seager:1999bc}. Based on the study of Ref.~\cite{Rubino-Martin_10}, which accounts for the non-equilibrium of angular momentum substates, the current version of \recfast~uses an updated fudge factor $F \approx 1.125$. It was shown in \cite{AliHaimoud:2010ab} that $\mathcal{A}_B/\alpha_B$ lies indeed in the range $1.12-1.14$, though it is not a constant but depends on redshift.  

The latest version of \recfast~corrects the net decay rate in Lyman-$\alpha$ which is in our equation \eqref{eq:Delta} (note, however, that the base model is different). The function $\Delta(z)$ is a sum of two Gaussians, whose amplitudes and widths were chosen to best mimick detailed calculations of \hyrec~and \textsc{cosmorec}.

It should be clear that the three-level simplification (and especially the fudging of $\mathcal{A}_B$) \emph{does not provide any computational advantage} over the already very simple effective 4-level model on which \hyrec-2~is based.

\end{appendix}

\bibliography{mybib}

\end{document}